\documentclass[11pt]{article}
\usepackage[font=footnotesize]{caption}

\usepackage{amsmath, amsfonts, amsthm, amssymb, amsrefs} 
\usepackage{mathtools}
\usepackage{tikz, graphicx, pgfplots, color, xcolor}
	\pgfplotsset{compat=1.12} 
	\usetikzlibrary{positioning}
\usepackage{enumitem}
\usepackage[normalem]{ulem}	

\usepackage[left=1.2in,right=1.2in,top=1in,foot=1in]{geometry}
\usepackage[linkbordercolor=cyan, pdfborder={0 0 0.5}]{hyperref}	

\newcommand{\benum}[2]{
	\begin{enumerate}[
		leftmargin={#2}, label={#1} 
		]
	}

\usepackage{amsthm}
	\newtheorem{thm}{Theorem}[section]

	\theoremstyle{definition}
		\newtheorem{defn}[thm]{Definition}
		\newtheorem{ex}[thm]{Example}
	\newtheorem{rmk}[thm]{Remark}

	\newcommand{\df}{\bf}		
	\let\originaleqref=\eqref 
		\renewcommand{\eqref}{Eq.~\originaleqref}

\newcommand{\eq}[1]{\begin{align*}#1\end{align*}}
\newcommand{\eqn}[1]{\begin{align}#1\end{align}}
\newcommand{\ds}{\displaystyle}			
\renewcommand{\emptyset}{\varnothing}	
\renewcommand{\epsilon}{\varepsilon}	
\renewcommand{\phi}{\varphi}			

\definecolor{orange}{RGB}{250, 140, 0}
	
\definecolor{turq}{RGB}{0, 160, 160}

\newcommand{\rr}{\ensuremath{\mathbb{R}}}

\usetikzlibrary{cd, arrows, automata}
	\newcommand*{\arrow}[1]{\arrow[#1]}

\usepackage{chemfig}

\DeclareMathOperator{\Span}{span}		
		\DeclareFontFamily{U}{mathx}{\hyphenchar\font45}
		\DeclareFontShape{U}{mathx}{m}{n}{
	      <5> <6> <7> <8> <9> <10>
	      <10.95> <12> <14.4> <17.28> <20.74> <24.88>
	      mathx10
	      }{}
		\DeclareSymbolFont{mathx}{U}{mathx}{m}{n}
		\DeclareFontSubstitution{U}{mathx}{m}{n}
		\DeclareMathAccent{\widecheck}{0}{mathx}{"71}

\newcommand{\kk}{k}

\newcommand{\bb}[1]{{\boldsymbol{#1}}}

\newcommand{\rrp}{\rr_{\geq 0}}
\newcommand{\rrpp}{\rr_{>0}}
\newcommand{\RR}{\ensuremath{\rightleftharpoons}}
\newcommand{\FR}{\ensuremath{\rightarrow}}

\let\originaleqref=\eqref
\renewcommand{\eqref}{Eq.~\originaleqref}

\title{Mathematical Analysis of Chemical Reaction Systems}

\author{
Polly Y. Yu \thanks{Department of Mathematics, University of Wisconsin-Madison; partially supported by NSERC PGS-D and NSF grant 1412643.}
\and
   Gheorghe Craciun \thanks{Department of Mathematics and Department of Biomolecular Chemistry, University of Wisconsin-Madison; partially supported by NSF grant 1412643.} 
}

\newcommand\captionfontsize {\footnotesize }

\begin{document}

\maketitle 



{\bf Abstract.} The use of mathematical methods for the analysis of chemical reaction systems has a very long history, and involves many types of models: deterministic versus stochastic, continuous versus discrete, and homogeneous versus spatially distributed. Here we focus on mathematical models based on deterministic mass-action kinetics. These models are systems of coupled nonlinear differential equations on the positive orthant. We explain how mathematical properties of the solutions of mass-action systems are strongly related to  key properties of the networks of chemical reactions that generate them, such as specific versions of reversibility and feedback interactions.


\section{Introduction}
\label{sec:MAS}

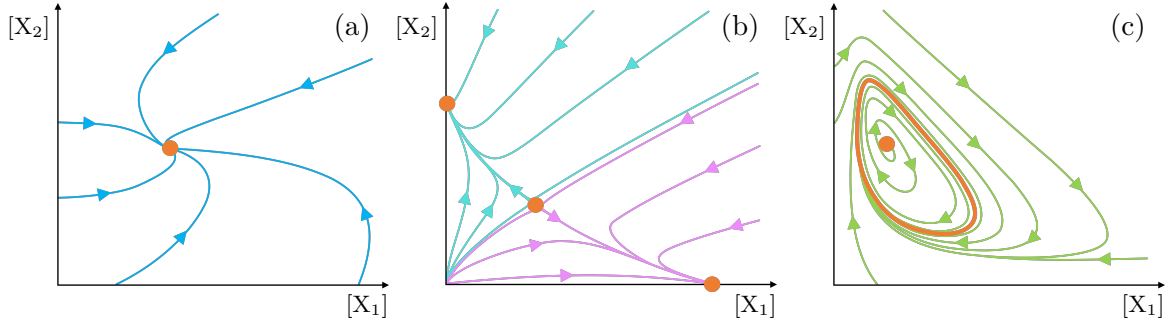
\begin{figure}[h!]
	\centering
	\includegraphics[width=1.8in]{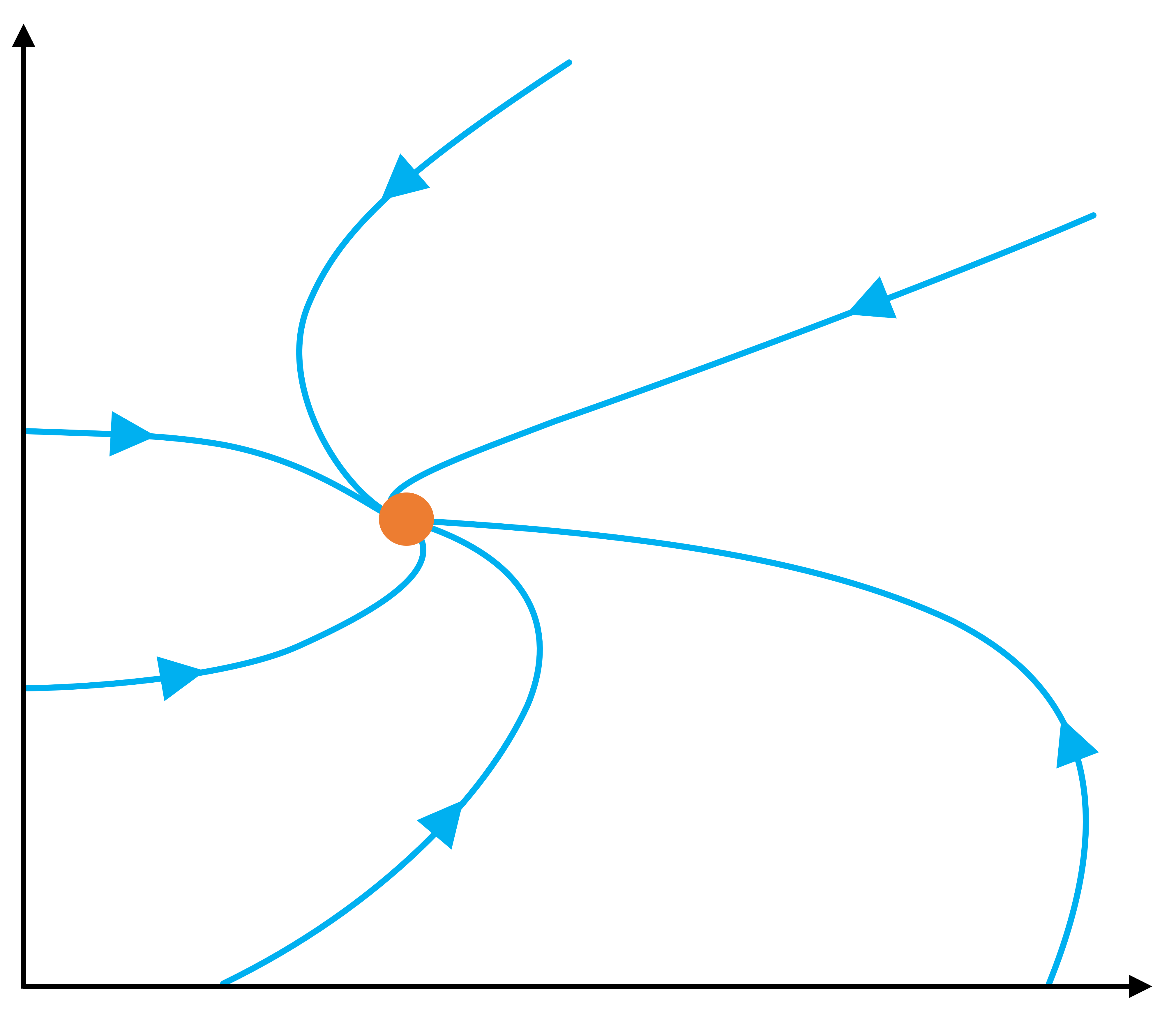}
	\begin{tikzpicture}[overlay]
		\node at (-0.7,3.5) {(a)};
		\node at (-5, 3.5) {\footnotesize [X${}_2$]};
		\node at (-0.5, -0.2) {\footnotesize [X${}_1$]};
	\end{tikzpicture}
	\hspace{0.2cm}
	\includegraphics[width=1.8in]{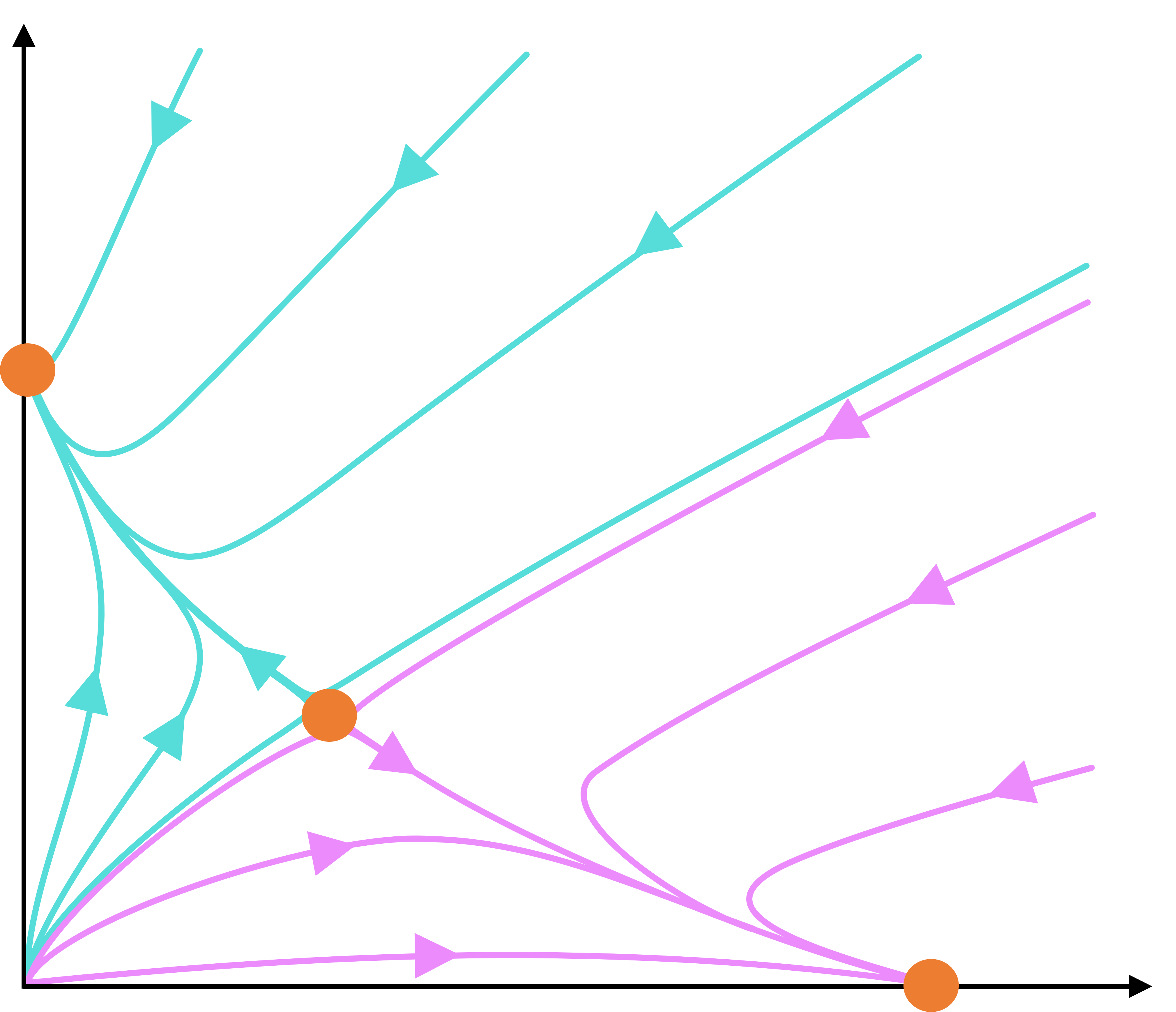}
	\begin{tikzpicture}[overlay]
		\node at (-0.7,3.5) {(b)};
		\node at (-5, 3.5) {\footnotesize [X${}_2$]};
		\node at (-0.5, -0.2) {\footnotesize [X${}_1$]};
	\end{tikzpicture}
	\hspace{0.2cm}
	\includegraphics[width=1.8in]{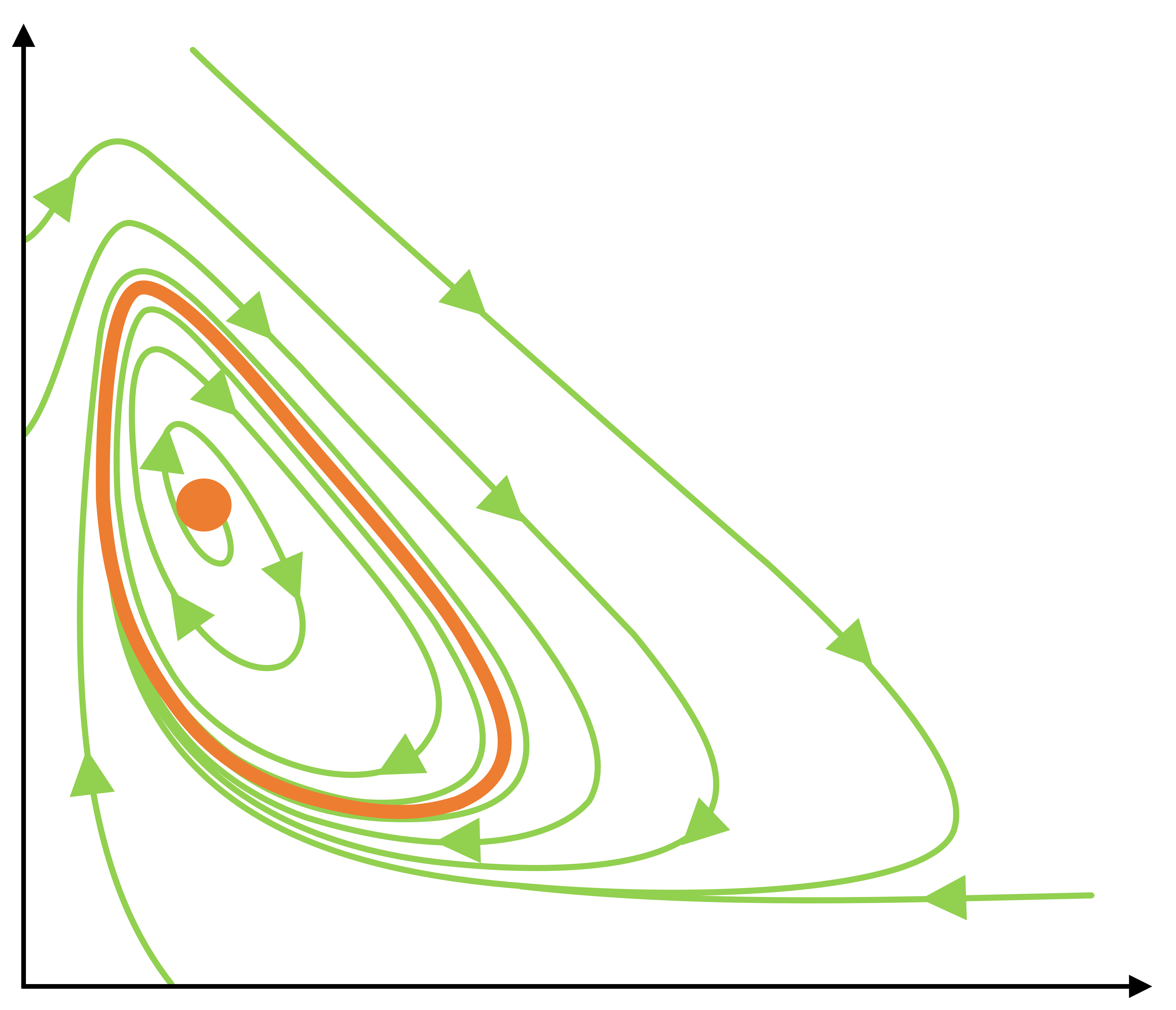}
	\begin{tikzpicture}[overlay]
		\node at (-0.7,3.5) {(c)};
		\node at (-5, 3.5) {\footnotesize [X${}_2$]};
		\node at (-0.5, -0.2) {\footnotesize [X${}_1$]};
	\end{tikzpicture}
	\hspace{0.75cm}
	\caption{\captionfontsize Phase portraits showing possible behaviors of mass-action systems: (a) uniqueness and stability of steady state, (b) bistability, and (c) oscillation. The mass-action systems, with the rate constants labeled on the reaction edges, are (a) \,  $\emptyset \xrightarrow{0.7} \rm{X}_1 \xrightarrow{\,1\,} \rm{X}_1 + \rm{X}_2 \xrightarrow{\,1\,} \emptyset$, \, (b) \, $\rm{X}_1 + \rm{X}_2 \xrightarrow{\,1\,} \rm{X}_1  \xrightleftharpoons[\,1\,]{3} 2 \rm{X}_1$ and $\rm{X}_1 + \rm{X}_2 \xrightarrow{\,2\,} \rm{X}_2  \xrightleftharpoons[\,1\,]{2} 2 \rm{X}_2$, \, (c) \, a version of the Selkov model, or ``Brusselator'', whose network (N2) is shown in Figure~\ref{fig:Selkovreactions} and rate constants given in Remark~\ref{rmk:oscillation}. }
	\label{fig:behaviors}
\end{figure}

Standard deterministic mass-action kinetics says that the rate at which a reaction occurs is directly proportional to the concentrations of the reactant species. For example, according to mass-action kinetics, the rate of the reaction $\rm{X}_1 + \rm{X}_2 \to \rm{X}_3$ is of the form $\kk x_1 x_2$, where $x_i$ is the concentration of species $\rm{X}_i$ and $\kk$ is a positive constant. If we are given a network that contains several reactions, then terms of this type can be added together to obtain a mass-action model for the whole network (see example below).
The law of mass-action was first formulated by Guldberg and Waage \cite{Guldberg_Waage_1864} and has recently celebrated its 150th anniversary \cite{Voit2015}. Mathematical models that use mass-action kinetics (or kinetics derived from the law of mass-action, such as Michaelis-Menten kinetics or Hill kinetics) are ubiquitous in chemistry and biology~\cite{Feinberg_1979, Clarke_1980, Alon, gunawardena, Voit2015, ErdiToth, Ingalls, Sontag1, deJong, Gilles}. The possible behaviors of mass-action systems also vary wildly; there are systems that have a single steady state for all choices of rate constants (Figure~\ref{fig:behaviors}(a)), systems that have multiple steady states (Figure~\ref{fig:behaviors}(b)), systems that oscillate (Figure~\ref{fig:behaviors}(c)), and systems (e.g. a version of the Lorentz system) that admit chaotic behavior~\cite{chaos}.

\begin{figure}[h!]
		\centering
		\includegraphics{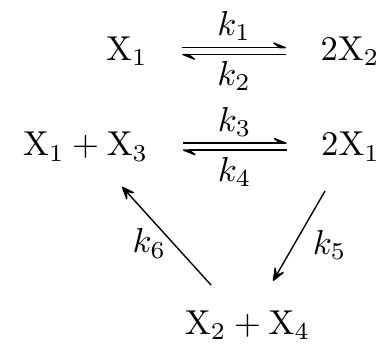}   		
		 \begin{tikzpicture}[overlay]
		 	\node at (5.1,2) {(N1)};
		 \end{tikzpicture}
		 \caption{\captionfontsize Example network (N1).}
		\label{fig:FigIntroExample}
	\end{figure}
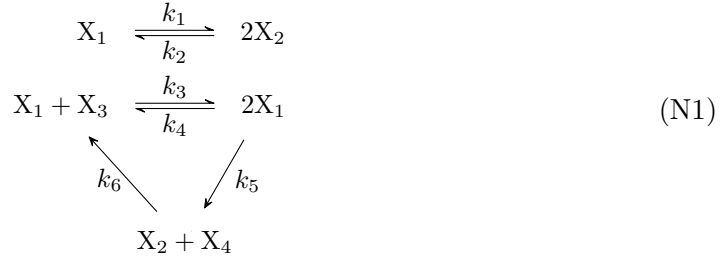

	To illustrate mass-action kinetics, consider the reaction network (N1) in Figure~\ref{fig:FigIntroExample}.
\noindent According to mass-action kinetics, the network (N1) gives rise to the following  system of differential equations on the positive orthant $\rrpp^4$:

	\eqn{
	\label{eq:introexample}
		\frac{dx_1}{dt} &= 
			-\kk_1 x_1 + \kk_2 x_2^2 + \kk_3 x_1 x_3 - \kk_4 x_1^2 -2 \kk_5 x_1^2 + \kk_6 x_2x_4 
			\nonumber \\
		\frac{dx_2}{dt} &= 
			2\kk_1 x_1 - 2\kk_2 x_2^2 + \kk_5 x_1^2 - \kk_6 x_2 x_4 
			 \\
		\frac{dx_3}{dt} &= 
			-\kk_3 x_1 x_3 + \kk_4 x_1^2 + \kk_6 x_2x_4 
			\nonumber \\
		 \frac{dx_4}{dt} &= 
		 	\kk_5 x_1^2 - \kk_6 x_2x_4,
		 	\nonumber
	}	
where $x_i = [\chemfig{X_i}]$ is the concentration of species $\chemfig{X_i}$.

	At any given time, the concentration vector $\bb x(t) = (x_1(t),x_2(t),\dots, x_n(t))^T$ is a point in $\rrpp^n$. Tracing the path over time gives a \emph{trajectory} in the \emph{state space} $\rrpp^n$. For example, Figure~\ref{fig:behaviors} shows several trajectories of three mass-action systems. For this reason, any concentration vector $\bb x = (x_1,x_2,\dots, x_n)^T$ is also called a {\df{state}} of the system, and we will refer to it as such. 

	\bigskip
	
\noindent 
In vector-based form, this dynamical system can also be written as

	\eqn{
	\label{eq:introexampleVec}
	\! \! \! \! \! \! \! \! \! \! \! \! \! \! \! \! \! \! \!
		\frac{d}{dt} \! \! 
			\begin{pmatrix} 
				x_1 \\ x_2 \\ x_3 \\ x_4 
			\end{pmatrix}
		\! 
		&=
		\kk_1 x_1  \! \! 
			\begin{pmatrix}
			-1 \\ 2 \\ 0 \\ 0 
			\end{pmatrix}\!\!
		+ \kk_2 x_2^2  \! \!
			\begin{pmatrix} 
			1 \\ -2 \\ 0 \\ 0
			\end{pmatrix}\!\!
		+ \kk_3 x_1 x_3  \! \!
			\begin{pmatrix}
			1 \\ 0 \\ -1 \\ 0
			\end{pmatrix}\!\!
		\nonumber \\
		& \qquad + \kk_4 x_1^2  \! \!
			\begin{pmatrix}
				-1 \\ 0 \\ 1 \\ 0
			\end{pmatrix}\!\!
		+ \kk_5 x_1^2  \! \!
			\begin{pmatrix}
			-2 \\ 1 \\ 0 \\1 
			\end{pmatrix}\!\!
		+ \kk_6 x_2x_4 \! \!
			\begin{pmatrix}
			1 \\ -1 \\ 1 \\ -1
			\end{pmatrix}	.		
	}

	\bigskip

	In order to write down a general mathematical formula for mass-action systems we need to introduce more definitions and notation.

	\bigskip

	The objects that are the source or the target of a reaction are called {\df{complexes}}. For example, the  complexes in the network (N1) are $ \rm{X}_1$, $2\rm{X}_2$, $\rm{X}_1+\rm{X}_3$, $2\rm{X}_1$, and $\rm{X}_2 + \rm{X}_4$. Their {\df{complex vectors}} are the vectors $\ds \begin{pmatrix} 1 \\ 0 \\ 0 \\ 0 \end{pmatrix}$, $\ds \begin{pmatrix} 0 \\ 2 \\ 0 \\ 0 \end{pmatrix}$,
	$\ds \begin{pmatrix} 1 \\ 0 \\ 1 \\ 0 \end{pmatrix}$,
	$\ds \begin{pmatrix} 2 \\ 0 \\ 0 \\ 0 \end{pmatrix}$,
	and 
	$\ds \begin{pmatrix} 0 \\ 1 \\ 0 \\ 1 \end{pmatrix}$, respectively.

	\bigskip

\noindent	Let us introduce the notation 
	\eqn{
		\label{eq:exponential}
		\bb x^\bb y = x_1^{y_1} x_2^{y_2} \cdots x_n^{y_n}
	} 
for any two vectors $\bb x = (x_1,x_2,\dots, x_n)^T \in \rr^n_{>0}$ and $\bb y = (y_1,y_2,\dots, y_n)^T \in \rr^n_{\ge0}$. Then the monomials $x_1$, $x_2^2$, $x_1x_3, \dots$, in the reaction rate functions in (\ref{eq:introexampleVec}) can be represented as $\bb x^\bb y$, where $\bb x$ is the  vector of  species concentrations and $\bb y$ is the complex vector of the source of the corresponding reaction. For example, $\rm{X}_1 + \rm{X}_3$ is the source of the reaction $\rm{X}_1 + \rm{X}_3 \to \rm{X}_4$, its complex vector  is $\ds \begin{pmatrix} 1 \\ 0 \\ 1 \\ 0 \end{pmatrix}$, and the corresponding reaction rate function in (\ref{eq:introexampleVec}) is $\kk_3 x_1x_3$.

	The vectors in (\ref{eq:introexampleVec}) are called {\df{reaction vectors}}, and they are the differences between the complex vectors of the target and source of each reaction; a reaction vector records the stoichiometry of the reaction. For example, the reaction vector corresponding to $\rm{X}_1 + \rm{X}_3 \to \rm{X}_4$ is $\ds \begin{pmatrix}
	-1 \\ 0 \\ -1 \\ 1 
	\end{pmatrix}
	=
	\begin{pmatrix}
	0 \\ 0 \\ 0 \\ 1 
	\end{pmatrix}
	-
	\begin{pmatrix}
	1 \\ 0 \\ 1 \\ 0 
	\end{pmatrix}
	$.
	
	
	There is a naturally defined oriented graph underlying a reaction network, namely the graph where the vertices are complexes, and the edges are reactions.	 Therefore, a chemical reaction network can be regarded as a {\df{Euclidean embedded graph}} $G = (V,E)$, where $V \subset \rr^n_{\ge 0}$ is the set of vertices of the graph, and $E \subset V \times V$ is the set of oriented edges of $G$. 
	
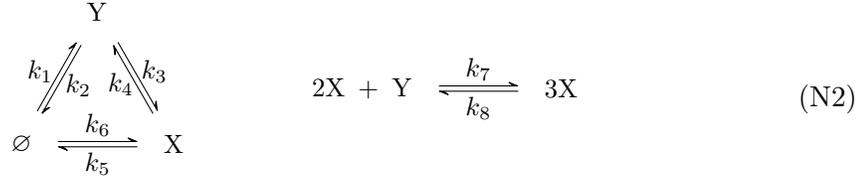
\begin{figure}[h!]
	\centering
		\includegraphics{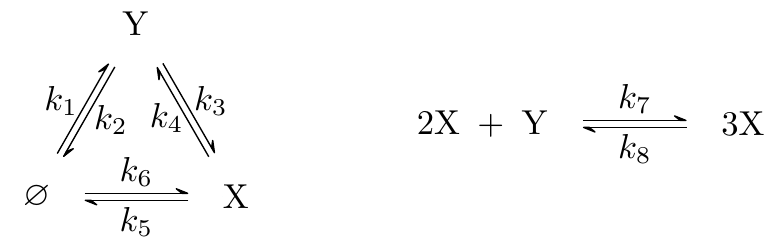} 
		
		 \begin{tikzpicture}[overlay]
		 	\node at (7.15,1.6) {(N2)};
		 \end{tikzpicture}
		 \caption{\captionfontsize  Example network (N2), a version of the Selkov model for glycolysis. It can also be regarded as a version of the ``Brusselator''.}
		\label{fig:Selkovreactions}
	\end{figure}

\begin{figure}[h!]
	\centering
		\includegraphics{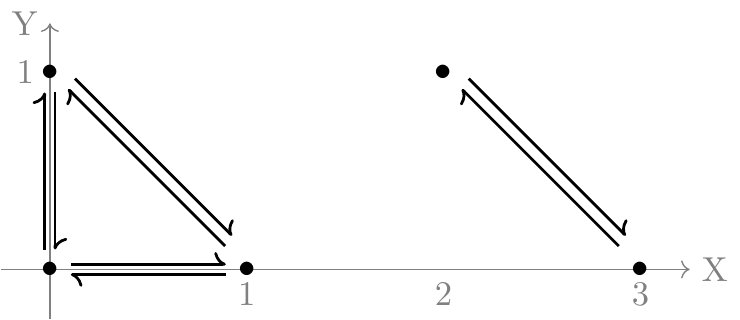}   	
		\caption{\captionfontsize  The Euclidean embedded graph $G$ of the network (N2) as shown in Figure~\ref{fig:Selkovreactions}.}
		\label{fig:SelkovEGraph}
	\end{figure}

	For example, (N2) depicted in Figure~\ref{fig:Selkovreactions} is the network for a version of the Selkov model for glycolysis.
Its Euclidean embedded graph $G$ in $\rrp^2$ is shown in Figure~\ref{fig:SelkovEGraph}.

	Given a reaction network with its Euclidean embedded graph $G$, and given a vector of reaction rate constants $\bb \kk$, we can use the notation (\ref{eq:exponential}) to write the {\bf mass-action system generated by $(G, \bb \kk)$} as shown in (4)
	\eqn{
	\label{eq:MAS}
		\frac{d\bb x}{dt} &= \sum_{ \bb y \to \bb y' \in G} \kk_{ \bb y \to \bb y'} \bb x^{\bb y} \left( \bb y' - \bb y \right).
	}

The {\df{stoichiometric subspace}} $S$ of a reaction network $G$ is the vector space spanned by its reaction vectors:
		\eqn{
			\label{eq:S}
			S = \Span_\rr \{ \bb y' - \bb y : \bb y \to \bb y' \in G\}.
		}
The {\df{stoichiometric compatibility class }} of $\bb x_0 \in \rrpp^n$ is the set $(\bb x_0 + S)_{>0} = (\bb x_0 + S)\cap \rrpp^n$, i.e., the intersection between the affine set $\bb x_0 + S$ and the positive orthant. Note that the solution $\bb x(t)$ of the mass-action system with initial condition $\bb x_0$ is confined to $(\bb x_0 + S)_{>0}$ for all future time, i.e., each stoichiometric compatibility class is a forward invariant set~\cite{Feinberg_1979}.
	
We say that a network or a graph $G$ is {\df{reversible}} if  $\bb y' \to \bb y$ is a reaction whenever $\bb y \to \bb y'$ is a reaction. We say that $G$ is {\df{weakly reversible}} if every reaction is part of an oriented cycle, i.e., each connected component of the graph $G$ is strongly connected. The network (N1) is weakly reversible, while (N2) is reversible. When the underlying graph $G$ is {{weakly reversible}}, we will see that the solutions of the mass-action system are known (or conjectured) to have many important properties, such as existence of positive steady states for all parameter values, persistence, permanence, and if the network satisfies some additional assumptions, also global stability~\cite{CNP, TDS, Craciun_GAC, Horn_Jackson, gunawardena, Feinberg_1979}. 
	
	For example, in the next section we will see that the mass-action systems generated by network (N1) and any values of rate constants are globally stable, i.e., there exists a globally attracting steady state within each stoichiometric compatibility class.

\section{Results Inspired by Thermodynamic Principles}
\label{sec:thermo}

	The idea of relating chemical kinetics and thermodynamics has a very long history, starting with Wegscheider~\cite{Wegscheider_1902}, and continuing with Lewis~\cite{Lewis_1925}, Onsager~\cite{Onsager_1931}, Wei and Prater~\cite{WeiPrater_1962}, Aris~\cite{Aris_1965}, Shear~\cite{Shear_1967}, Higgins~\cite{Higgins_1968}, and many others. For example, the notion of ``detailed-balanced systems" was studied in depth, and this notion has a strong connection to the thermodynamical properties of microscopic reversibility which goes back to Boltzmann~\cite{Boltzmann_1887, Boltzmann_1896, Gorban_Karlin_2005}.

\subsection{Detailed-Balanced and Complex-Balanced Systems}
\label{sec:DBCB}
	
	In 1972 Horn and Jackson~\cite{Horn_Jackson} have identified the class of ``complex-balanced systems" as a generalization of detailed-balanced systems. While complex-balanced systems are not necessarily thermodynamically closed systems, Horn and Jackson were interested in systems that behave as though the laws of thermodynamics for closed systems are obeyed. In particular, according to the Horn-Jackson theorem below, a complex-balanced system has a unique steady state within each stoichiometric compatibility class, and it is locally stable within it~\cite{Horn_Jackson}.

	Of all the positive steady states, we call attention to two kinds that are especially important. These are characterized by the \emph{fluxes} at a state $\bb x_0$, i.e., the values $\kk_{\bb y \to \bb y'} \bb x_0^{\bb y}$ of the reaction rate functions evaluated at $\bb x_0$.
	\begin{defn}
		A state $\bb x_0$ of a mass-action system is a {\df{detailed-balanced steady state}} if the network is reversible, and every forward flux is balanced by the backward flux at that state, i.e., for every reaction pair $\bb y \rightleftharpoons \bb y'$, we have
		\eqn{
		\label{eq:DB}
			\kk_{\bb y \to \bb y'} \bb x_0^{\bb y} = \kk_{ \bb y' \to  \bb y} \bb x_0^{\bb y'}.
		}
In particular, if a network is not reversible, then it cannot admit a detailed-balanced steady state. 
		
		A state $\bb x_0$ of a mass-action system is a {\df{complex-balanced steady state}} if at each vertex of the corresponding Euclidean embedded graph $G$, the fluxes flowing into the vertex balance the fluxes flowing out of the vertex at that state $\bb x_0$, i.e., for every complex $\bb y$ we have
		\eqn{
		\label{eq:CB}
			\sum_{ \bb y \to  \bb y' \in G} \kk_{ \bb y \to  \bb y'} \bb x_0^{\bb y} \ = \sum_{ \bb y' \to  \bb y \in G} \kk_{ \bb y' \to  \bb y} \bb x_0^{\bb y'} 
		}
In particular, it can be shown that if the network is not weakly reversible, then it cannot admit a complex-balanced steady state.
	\end{defn}
	
	At a detailed-balanced steady state $\bb x_0$, the fluxes across pairs of reversible reactions are balanced; hence $\bb x_0$ is also called an \emph{edge-balanced steady state}. At a complex-balanced steady state $\bb x_0$, the net flux through any vertex is zero; hence $\bb x_0$ is also called a \emph{vertex-balanced steady state}.

	\begin{defn}
		A {\df{detailed-balanced system}} is a mass-action system $(G,\bb \kk)$ that has at least one detailed-balanced steady state. 
		A {\df{complex-balanced system}} is a mass-action system $(G,\bb \kk)$ that has at least one complex-balanced steady state.
	\end{defn}

	It is not difficult to check that if the state $\bb x_0$ is detailed-balanced, then it is complex-balanced, i.e., complex balance is a generalization of detailed balance. Complex-balanced systems enjoy many properties of detailed-balanced systems; the Horn-Jackson theorem is the first such result.

	\begin{thm}[{\bf Horn-Jackson theorem}~\cite{Horn_Jackson}]
	Consider a reaction network $G$ and a vector of reaction rate constants $\bb \kk$. Assume that the mass-action system generated by $(G, \bb \kk)$ has a complex-balanced steady state $\bb x^*$; in other words, $(G,\bb \kk)$ is a complex-balanced system.
		Then all of the following properties hold:
		\begin{enumerate}
		\item
			All positive steady states are complex-balanced, and there is exactly one steady state within every stoichiometric compatibility class.
		\item
			The set of complex-balanced steady states $Z_{\bb \kk}$ satisfies the equation $\ln Z_{\bb \kk} = \ln \bb x^* + S^\perp$, where  $S$ is the stoichiometric subspace of $G$.
		\item
			The function
			\eqn{
				\label{eq:Lyapunov}
				L(\bb x) = \sum_{j=1}^n x_i (\ln x_i - \ln x_i^* - 1).
			}
			is a strictly convex Lyapunov function of this system, defined on $\rrpp^n$ and with global minimum at $\bb x = \bb x^*$.
		\item
			Every positive steady state is locally asymptotically stable within its stoichiometric compatibility class.
		\end{enumerate}
	\end{thm}

	\rmk
	The original paper of Horn and Jackson~\cite{Horn_Jackson} claimed that each complex-balanced steady state is a \emph{global} attractor within its stoichiometric compatibility class. Later, Horn~\cite{Horn_1974} realized that this claim does not follow from the existence of the Lyapunov function above, and formulated it as a conjecture, later known as the ``global attractor conjecture'' (see Section~\ref{sec:GAC}). 
	
	\rmk 
	\label{rmk:oscillation}
	Horn and Jackson referred to the Lyapunov function (\ref{eq:Lyapunov}) as a ``pseudo-Helmholtz function''. This function can be regarded as a finite-dimensional version of the Boltzmann entropy, and the fact that it decreases along trajectories of  a complex-balanced system can be regarded as a version of the Boltzmann's H-theorem~\cite{Gorban_Karlin_2005}. Shear~\cite{Shear_1967} claimed this to be true for any steady state of a reversible system (not necessarily complex-balanced), but this claim was later shown to be false by Higgins~\cite{Higgins_1968}. For example, the mass-action system generated by network (N2) in Figure~\ref{fig:Selkovreactions} with rate constants 	
	\eq{
		\kk_1 = 0.5, \quad 
		\kk_2 = \kk_6 = \kk_8 = 0.1, \quad 
		\kk_3 = \kk_4 = 0.01, \quad 
		\kk_5 = \kk_7 = 1,
	} 
has a unique positive steady state that is unstable and sits inside a stable limit cycle. Several trajectories, including the limit cycle, of this mass-action system are featured in Figure~\ref{fig:behaviors}(c). 
	
\rmk
\label{rmk:BEC}
As we discussed earlier, ideas inspired by thermodynamics (specifically the Boltzmann equation) can be used to analyze chemical reaction networks. On the other hand, results obtained for chemical reaction networks can be used to analyze discrete versions of the Boltzmann equation~\cite{Craciun_Tran}.
	
\subsection{Deficiency Theory}
\label{sec:deficiency}
	
	The existence of a complex-balanced steady state is difficult to check in practice, but simple sufficient conditions for complex balance exist. The best known result, due to Feinberg and Horn~\cite{Horn_1972, mf72, HornFeinberg74}, is based on \emph{deficiency}, a non-negative integer associated to a reaction network.
 
\begin{defn}
If the underlying graph $G$ of a reaction network has $m$ nodes and $\ell$ connected components, and the dimension of the stoichiometric subspace is $s$, then the {\df{deficiency}} of the network is the non-negative integer $\delta = m - \ell - s$. 
\end{defn}

\begin{thm}[{\bf Deficiency zero theorem}~\cite{Horn_1972, mf72, HornFeinberg74}]
		A mass-action system is complex-balanced for all values of its reaction rate constants if and only if it is weakly reversible and has deficiency zero. 
\end{thm}

The complex balance property has rich algebraic structure, and deficiency can be regarded as a measure of how far a weakly reversible system is from being complex-balanced. In particular it has been shown that, given a weakly reversible network $G$, the mass-action system generated by $(G,\bb \kk)$ is complex-balanced if and only if the vector of reaction rate constants $\bb \kk$ lies on an algebraic subvariety of codimension $\delta$~\cite{TDS}. In this context, the deficiency zero theorem refers to the codimension zero case, i.e., the case where the mass-action system is complex-balanced for all $\bb \kk$. 

\begin{ex}
Consider again the reaction network (N1) from Figure~\ref{fig:FigIntroExample}. This network is weakly reversible and has deficiency $\delta = 5 - 2 -3 = 0$. Therefore, according to the deficiency zero theorem, the network (N1) is complex-balanced for all values of its reaction rate constants $\kk_1, \kk_2, \dots, \kk_6$. Furthermore, according to the Horn-Jackson theorem, it follows that the mass-action system (2) has a unique (locally asymptotically stable) steady state within each stoichiometric compatibility class, for all choices of reaction rate constants. Global stability follows from recent results in~\cite{Craciun_GAC}.
\end{ex}

\begin{ex}
\label{ex:oscillation}
The dynamical properties of mass-action systems with $\delta > 0$ may depend on the values of the rate constants $\kk_i$. For example, the network (N2) in Figure~\ref{fig:Selkovreactions} has deficiency $\delta = 5-2-2 = 1$. The deficiency zero theorem is silent in this case. Indeed, we have already seen in Remark~\ref{rmk:oscillation} that for some chosen rate constants, this system has a limit cycle and an unstable steady state. However, if we choose all $\kk_i = 1$, the system is complex-balanced, and thus it has a unique locally asymptotically stable steady state. 
\end{ex}

\rmk
	If a network has deficiency $\delta = 0$ but is not weakly reversible, then it cannot have any positive steady states, i.e., its steady states (if there are any) must be on the boundary of $\rrp^n$~\cite{Feinberg_1979}. On the other hand, according to Feinberg's ``deficiency one theorem", some networks that have $\delta > 0$ are known to have a unique steady state within each stoichiometric compatibility class for all rate constants~\cite{Feinberg_1979, Feinberg_1987, Feinberg_1995, Boros}. For results on existence of steady states for ``generalized mass-action systems'', see~\cite{Muller_Regensburger_2012, genMAS18}.

\section{Multistability and Chemical Switches}
\label{sec:multistable}

There is great interest in biological  applications in understanding ``biochemical switches'', i.e., reaction networks that have multiple positive steady states within the same stoichiometric compatibility class. As a consequence of the Horn-Jackson theorem, if the network is weakly reversible, then these steady states \emph{cannot} be complex-balanced, and the deficiency of such a network must be strictly positive.

\subsection{The SR Graph}

Some mathematical criteria for multistability are able to detect very subtle differences between networks. One such approach was introduced in \cite{cf06} and is based on a bipartite labeled graph associated to the reaction network, called the {\df{species-reaction graph (SR graph)}}. 
	
	The SR graph is defined as follows. The nodes of the SR graph are either \emph{species nodes} (one for each chemical species in the network) or \emph{reaction nodes} (one for each reversible or irreversible reaction in the network). There are no edges between two species nodes, or between two reaction nodes. 
Consider a species node $\rm{X}$ and a reaction node $\bb y \to \bb y'$ (or $\bb y \rightleftharpoons \bb y'$). The SR graph contains an \emph{edge} between these two nodes if and only if  $\rm{X}$ is involved in this reaction, either as a reactant or as a product. 
Each edge has a \emph{complex label}, as follows. 
If $\rm{X}$ is a reactant (i.e., is in the support of $\bb y$), we label the edge with the complex $\bb y$; similarly, if $\rm{X}$ is a product (i.e., is in the support of $\bb y'$), we label the edge with the complex $\bb y'$. If $\rm{X}$ is {\em both} a reactant and a product,  we draw {\em two} edges from the species node to the reaction node, and we label one with $\bb y$ and the other with $\bb y'$.

\begin{figure}[h!]
		\centering
		\includegraphics{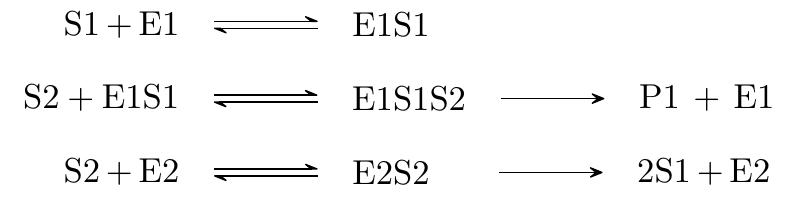}   		
		 \begin{tikzpicture}[overlay]
		 	\node at (3.1,1) {(N3)};
		 \end{tikzpicture}
		 \caption{\captionfontsize Example network (N3).}
		\label{fig:FigSRExample}
	\end{figure}
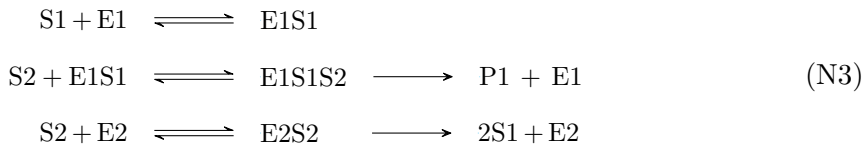

\begin{figure}[h!]
		\centering
		\includegraphics
			{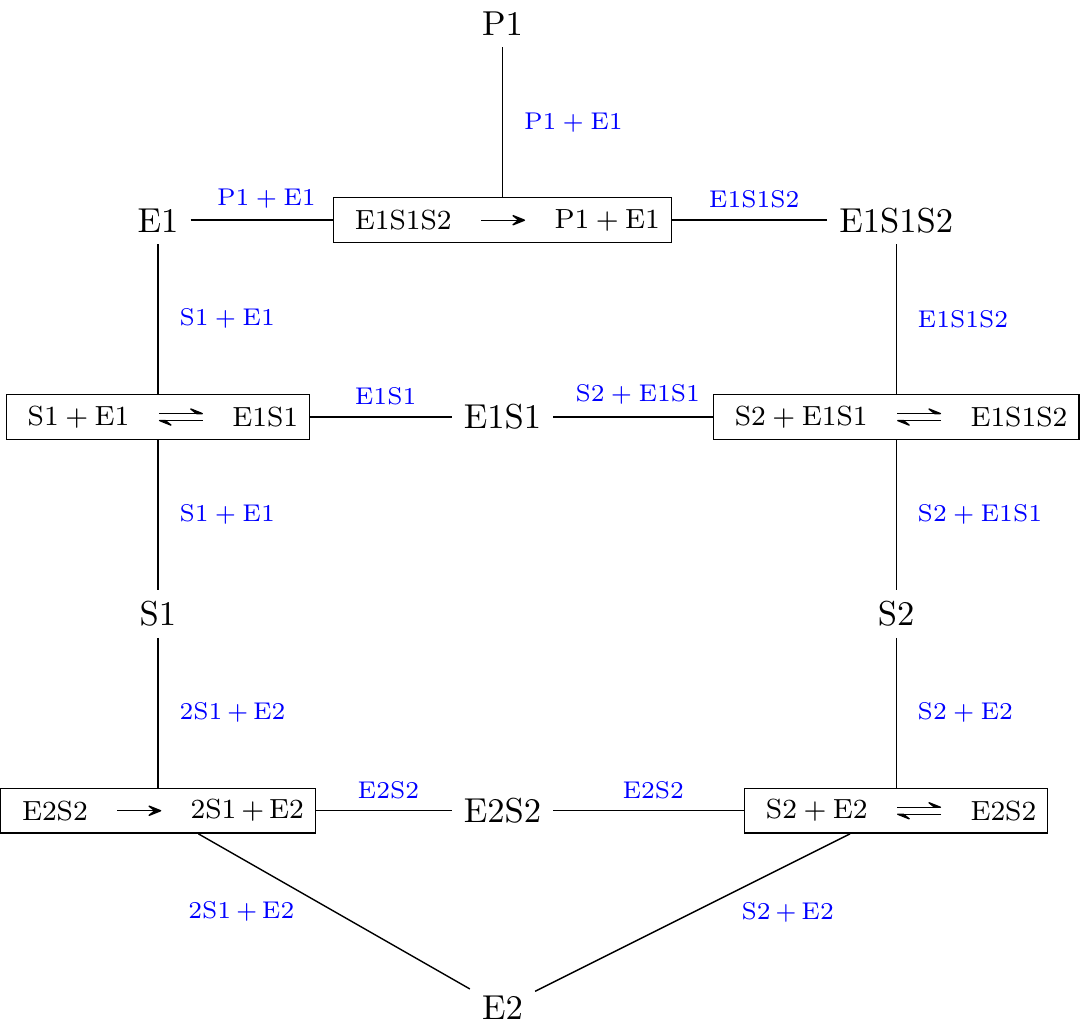}
		\caption{\captionfontsize The SR graph of  reaction network (N3) in Figure~\ref{fig:FigSRExample}.}
		\label{fig:figSRExample-SR}
	\end{figure}

\begin{ex}
	Consider the network (N3) in Figure~\ref{fig:FigSRExample}.
\end{ex}
\noindent
The SR graph of this network is shown in Figure~\ref{fig:figSRExample-SR}. Note how, for reversible reactions, the forward and backward reactions share the same reaction node in the SR graph. The complex labels of all the edges are shown in blue in Figure~\ref{fig:figSRExample-SR}. 

	In order to describe criteria for multistability that are based on the SR graph, we have to distinguish between various types of cycles that may occur in it.
	
\begin{defn}
	If a pair of edges in an SR graph shares a reaction node and have the same complex label, then it is called a {\df{c-pair}}. If a cycle in an SR graph contains an odd number of c-pairs, then it called an {\df{odd cycle}}; otherwise it is called an {\df{even cycle}}. The {\em stoichiometric coefficient of an edge} that is adjacent to species $\rm{X}$ and has complex label $\bb y$ is the coefficient of $\rm{X}$ within $\bb y$. If all the edges of a cycle have  stoichiometric coefficient equal to $1$, then that cycle is called a {\df{1-cycle}}. Also, we say that two cycles  {\df{have an S-to-R intersection}} if all connected components of their intersection are paths from a species node to a reaction node. 
\end{defn}

\newpage 

Using this classification of cycles, we can formulate the following necessary condition for multistability:
\begin{thm}[\cite{cf06, ctf06, cf10}]
\label{thm_SR}
	Assume that the SR graph of a reaction network satisfies the following two conditions:

		(i) all cycles are odd cycles or 1-cycles, and 
		
		(ii) no two even cycles have an {{S-to-R intersection}}.
		
\noindent Then the corresponding mass-action system cannot have multiple non-degenerate steady states within the same stoichiometric compatibility class, for any values of the reaction rate constants.
\end{thm}

Let us use this theorem to analyze the reaction network whose SR graph is shown in Figure~\ref{fig:figSRExample-SR}. Note that the stoichiometric coefficients of all the edges are 1, \emph{except} for the edge that connects the species $\rm{S}1$ and the reaction $\rm{E}2\rm{S}2 \to 2\rm{S}1+\rm{E}2$. Therefore, all the cycles are 1-cycles, except for the cycles that contain this particular edge. On the other hand, it is easy to check that the four cycles that contain this edge are odd cycles, so condition (\emph{i}) of the theorem above is satisfied. Condition (\emph{ii}) is also satisfied, since in order for two cycles to have an S-to-R intersection we would need to have at least one species node with three or more adjacent edges; but all species nodes in this SR graph have at most two adjacent edges.

The network (N3) contains two substrates $\rm{S}1$, $\rm{S}2$, and a single product $\rm{P}1$. A similar network with three substrates $\rm{S}1$, $\rm{S}2$, $\rm{S}3$, and two products $\rm{P}1$, $\rm{P}2$ {\em does} give rise to multistable systems for some values of the rate constants; in that case, some of the cycles that fail to be 1-cycles also fail to be even cycles, so condition (\emph{i}) of the theorem does not hold. On the other hand, a similar network with four substrates and three products  {\em cannot} give rise to multiple steady states, and so on.\cite{ctf06} This shows that the capacity for multistability is not only a result of having a complex network with many species and reactions, but more subtle features must be present, some of which are described by Theorem \ref{thm_SR}. 
More examples of the use of this theorem and related results can be found in \cite{cf06, ctf06}, and further results and generalizations can be found in \cite{Banaji_Craciun_2009, cf10}.

\subsection{The Jacobian Criterion}

The results presented in the previous section on the lack of multistability rely on the {\em injectivity} property for reaction networks (see below). This property was introduced in~\cite{cf05}, where the {\em Jacobian criterion} was shown to be a sufficient condition for injectivity, which, in turn, implies uniqueness of steady states for ``open" mass-action systems, i.e.,  systems where there is a non-negative inflow rate for each species, and also a positive outflow/degradation rate for each species. The inflow terms are represented by ``inflow reactions" of the form $\emptyset \to \rm{X}$, and the outflow/degradation terms are represented by ``outflow reactions" of the form $\rm{X} \to \emptyset$. Note that such a network has a single  stoichiometric compatibility class, which is the whole positive orthant. 

\begin{defn}
We say that a reaction network $G$ is {\bf injective} if the right-hand side of the differential equation (\ref{eq:MAS}), regarded as a function of  $\bb x$ is  injective (i.e., one-to-one)  for all values of the reaction rate constants.
\end{defn}

It is easy to see that injectivity implies that there cannot exist multiple steady states. In general it is difficult to check the global injectivity of nonlinear functions such as the right-hand side of (\ref{eq:MAS}). The following theorem (called the Jacobian criterion) addresses this challenge.

\begin{thm}[\cite{cf05, cf10}]
\label{thm:Jac}
Consider an open reaction network $G$. Then the following hold:
\begin{enumerate}
\item
	The open reaction network $G$ is injective if and only if the determinant of the Jacobian matrix of the right-hand side of its differential equation {\rm (\ref{eq:MAS})} is different from zero for all values of $\bb x$ and for all values of the reaction rate constants $\bb \kk$.
\item
	Consider a reaction network $G'$ such that its corresponding open reaction network is $G$, i.e., $G$ includes the reactions of $G'$ and inflow and outflow reactions for all species. If $G$ is injective, then for all choices of reaction rate constants, $G$ cannot give rise to multiple steady states, and $G'$ cannot give rise to multiple non-degenerate steady states. 
\end{enumerate} 
\end{thm}

\begin{ex}
\label{ex:jac}
For example, consider the network (N4) in Figure~\ref{fig:FigJacobianExample}.

	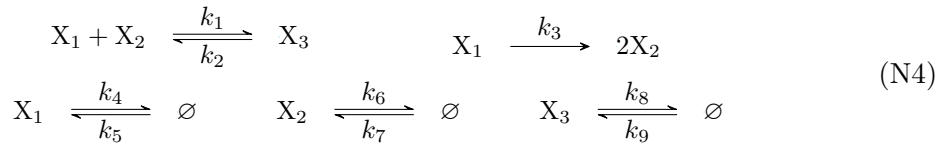
\begin{figure}[h!]
		\centering
		\includegraphics{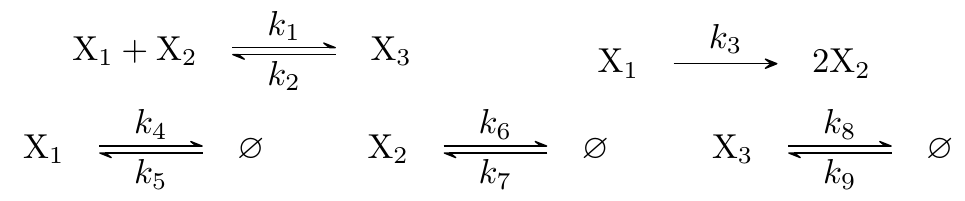}    		
		 \begin{tikzpicture}[overlay]
		 	\node at (2.2,1) {(N4)};
		 \end{tikzpicture}
		 \caption{\captionfontsize Example network (N4).}
		\label{fig:FigJacobianExample}
	\end{figure}
\end{ex}
\noindent
The reactions $\rm{X}_i\RR \emptyset$ are due to the inflow and outflow of the species $\rm{X}_i$. 
The mass-action system for this network is given by
	\eqn{
		\label{eq:JacobianMAS}
		\frac{dx_1}{dt} &=
			\kk_5 - \kk_4 x_1 
			- \kk_1 x_1x_2 + \kk_2 x_3
			- \kk_3 x_1 
			\nonumber \\
		\frac{dx_2}{dt} &=
			\kk_7 - \kk_6 x_2 
			- \kk_1 x_1x_2 + \kk_2 x_3 
			+ 2 \kk_3 x_1 
			 \\
		\frac{dx_3}{dt} &=
			\kk_9 - \kk_8 x_3 
			+ \kk_1 x_1x_2 - \kk_2 x_3.
			\nonumber
	}

\noindent
The Jacobian matrix of the right-hand side of (\ref{eq:JacobianMAS}) is 
	\eqn{
	\mathrm{Jac}(\bb x, \bb \kk) 
	= \begin{pmatrix}
		-\kk_4 - \kk_1 x_2 - \kk_3
		&
		-\kk_1 x_1 
		&
		\kk_2 
		\\ 
		-\kk_1 x_2 + 2 \kk_3
		&
		-\kk_6 - \kk_1 x_1 
		&
		\kk_2
		\\ 
		\kk_1 x_2
		&
		\kk_1 x_1 
		&
		-\kk_8 - \kk_2
	\end{pmatrix}.
	}

\noindent	
Then a simple calculation shows that we have
	\eqn{
		\det(\mathrm{Jac}(\bb x, \bb \kk))
		&=
		- \kk_4 \kk_6 \kk_8
		- \kk_2 \kk_4 \kk_6
		- \kk_1 \kk_4 \kk_8 x_1
		- \kk_1 \kk_6 \kk_8 x_2 
		\nonumber \\ 
		& \qquad
		-  \kk_3 \kk_6 \kk_8
		- \kk_2 \kk_3 \kk_6
		- 3 \kk_1 \kk_3 \kk_8 x_1.
	}
\noindent	
Since $\bb x$ and $\bb \kk$ have positive coordinates, this implies that the determinant of the Jacobian of the right-hand side of (\ref{eq:JacobianMAS}) is different from zero for all $\bb x$ and all $\bb \kk$.

Then, by applying Theorem \ref{thm:Jac}, it follows that the network (N4) cannot give rise to multiple steady states for any values of the reaction rate constants.

More details and examples about the use of the Jacobian criterion can be found in \cite{cf05}. Further results and generalizations for closed or ``semi-open" systems can be found in \cite{Banaji_Donnell_Baigent, Banaji_Craciun_2009, cf10, sf11, Feliu_Wiuf_2012, BadalAnne2015, BadalAnne2011, SignCondGenPolyMaps, Banaji_Pantea_2016}.

\section{Persistence and Global Stability}
\label{sec:GAC}

Starting with the work of Horn~\cite{Horn_1974, Horn_Jackson} and Feinberg~\cite{Feinberg_1987}, there has been ever an increasing interest in understanding the long-time dynamics of solutions of mass-action systems. For example, in 1974 Horn has conjectured that the unique complex-balanced steady state is not only locally stable, but is actually globally stable~\cite{Horn_1974}. This statement was later~\cite{TDS} called the {\em global attractor conjecture}:

\medskip

\noindent
{\bf Global Attractor Conjecture.} {\em Any complex-balanced mass-action system has a globally attracting point within every stoichiometric compatibility class. }

\medskip

\noindent This conjecture is widely regarded as the most important open problem in this field. Its study led to an increased interest in the limit behavior of solutions of mass-action systems as $t \to \infty$, and has inspired the following more general conjectures \cite{CNP}:

\medskip

\noindent
{\bf  Persistence Conjecture.}  {\em Any weakly reversible mass-action system is {\bf persistent}, i.e., its solutions cannot have a limit point on the boundary of the positive orthant.} 

\medskip

\noindent
{\bf  Permanence Conjecture.}  {\em Any weakly reversible mass-action system is {\bf permanent}, i.e., there exists a globally attracting compact set within every stoichiometric compatibility class.}

\medskip

Figure~\ref{fig:behaviors}(a) and (c) demonstrate persistence (i.e., none of the species become extinct) as well as permanence (i.e., the eventual concentrations of all the species become bounded and bounded away from 0). In contrast, the system corresponding to Figure~\ref{fig:behaviors}(b) is not persistent.

The global attractor conjecture is the oldest and best known of these conjectures, and has resisted efforts for a proof for over four decades, but proofs of many special cases have been obtained during this time, for example in~\cite{siegel_maclean, Sontag1, TDS, ShiuSturmfels, Anderson_2011, CNP, persistence2, angeli_deleenheer_sontag_2011}.
The conjecture originated from the 1972 breakthrough work by  Horn and Jackson~\cite{Horn_Jackson}, and was  formulated in its current  form by Horn in 1974~\cite{Horn_1974}.
As a historical note, actually Horn and Jackson~\cite{Horn_Jackson} stated that the complex-balanced steady state within each stoichiometric compatibility class is a {\em global attractor}, but soon afterwards Horn explained that they have {\em not} actually proved this claim, and he proposed this global convergence property as a conjecture~\cite{Horn_1974}.

Recently, Craciun, Nazarov and Pantea~\cite{CNP} have proved the three-dimensional case of the global attractor conjecture, and Pantea~\cite{persistence2} has generalized this result for the case where the dimension of the stoichiometric compatibility class is at most three. Using a different approach, Anderson~\cite{Anderson_2011} has proved the conjecture under the additional hypothesis that the reaction network has a single linkage class, and this result has been generalized by Gopalkrishnan, Miller, and Shiu~\cite{Gopalkrishnan_Miller_Shiu_2013} for the case where the reaction network is strongly endotactic. A proof of the global attractor conjecture {\em in full generality} has been recently proposed in~\cite{Craciun_GAC}.

	For example, the weakly reversible mass-action system with deficiency $\delta = 0$ corresponding to Figure~\ref{fig:behaviors}(a) has a globally attracting point for all choices of rate constants. This follows from the results in~\cite{CNP} because the system has dimension $n \leq 3$. Alternatively it also follows from~\cite{Anderson_2011} because it has a single linkage class. However, the reversible mass-action system corresponding to Figure~\ref{fig:behaviors}(c) is complex-balanced only for some choices of rate constants~\cite{TDS}, and therefore has a globally attracting point for those choices of rate constants~\cite{CNP}. In contrast, the mass-action system corresponding to Figure~\ref{fig:behaviors}(b) is never complex-balanced for any value of the rate constants, because it is not weakly reversible.

\medskip

The persistence conjecture and the permanence conjecture have only been proved for two-dimensional systems~\cite{CNP}. 

\medskip

\rmk
Note  that the applicability of all the results described in the previous sections can be extended by using the fact that a reaction network {\em is not  uniquely identified by the mass-action systems that it generates}. In other words, different networks may give rise to the same dynamical systems~\cite{Craciun_Pantea_2008}. Therefore, one may be able to deduce properties of the dynamical systems generated by a given network by using the fact that there exists \emph{another network} that gives rise to the same model, and this second network may exhibit useful properties that the first one did not (for example, the second network may be weakly reversible, or its SR graph may have useful properties~\cite{Craciun_Pantea_2008, Johnston_2012}).

\section{Other Models: Non-polynomial Models,  Stochastic Mass-Action Systems, Reaction-Diffusion Equations}
\label{sec:OtherModels}

We have mostly focused on deterministic mass-action kinetics, which is a finite-dimensional system of differential equations that is best suited for high molecular counts in well-mixed dilute solutions. Many other models of chemical reaction systems exist, and they may be more suitable for very low molecular counts and/or spatially inhomogeneous systems. In this section, we describe some of these types of models, all of which are active areas of research. 



\subsection{Anomalous Reaction Orders and Time-Dependent Reaction Rate Constants}
\label{sec:powerlaw}

	To model spatially inhomogeneous systems, one can modify mass-action kinetics in several ways. For example, one can allow the rate constants to become time-dependent~\cite{Kopelman, SchnellTurner}, or one can allow the kinetic orders (i.e., the powers in the reaction rate functions) to be different from standard mass-action kinetics~\cite{Kopelman, Savageau}. The latter is sometimes  called power-law kinetics. 

	In the case where the rate constants are time-dependent,  the system of differential equations becomes non-autonomous, but one may still be able to draw conclusions about persistence or permanence properties. One approach embeds this non-autonomous dynamical system into an autonomous differential inclusion model, and then  constructs forward invariant sets, i.e., invariant regions, for the  differential inclusion model~\cite{Brunner_Craciun, Craciun_GAC}.

	The reaction rate functions for power-law kinetics are generalized monomials (whose exponents may be non-integer). Some of the techniques used for the analysis of classical mass-action systems can be carried over to this setting~\cite{Horn_Jackson, CNP, Muller_Regensburger_2012, genMAS18}.

\subsection{Michaelis-Menten Kinetics, Hill Kinetics, and Quasi-Steady State Approximation}
\label{sec:rational}

	In biochemistry, it is common to see the Michaelis-Menten enzyme kinetics or the Hill binding kinetics. These are derived from mass-action kinetics by  \emph{quasi-steady state approximation}, which is a method of model reduction based on elimination of fast intermediates~\cite{ErdiToth, Ingalls, Craciun_Pantea_Rawlings}. 
	
	Mathematically, the Michaelis-Menten and the Hill kinetics give rise to reaction rate functions that are rational functions, i.e., ratios of polynomial functions. The analysis of these systems can be reduced to the analysis of dynamically equivalent mass-action systems by using time-rescaling to eliminate all denominators~\cite{Brunner_Craciun}.

	For example, consider the reversible reaction $\rm{X}_1 + \rm{X}_2 \RR 2\rm{X}_1$, where the forward reaction $\rm{X}_1 + \rm{X}_2 \to 2\rm{X}_1$ is modeled with a Michaelis-Menten reaction rate function $\ds \frac{\kk_1 x_1x_2}{\kk_2 + x_1}$, and the backward reaction $2\rm{X}_1 \to \rm{X}_1 + \rm{X}_2$ is modeled using standard mass-action kinetics with reaction rate function $\kk_3x_1^2$. Then the system of differential equations corresponding to these two reactions is
	\eq{
		\frac{dx_1}{dt} &= \frac{\kk_1 x_1x_2}{\kk_2 + x_1} - \kk_3 x_1^2  \\
		\frac{dx_2}{dt} &= -  \frac{\kk_1 x_1x_2}{\kk_2 + x_1} + \kk_3 x_1^2.
	}	
Instead of studying the above equations, one may study the mass-action system
	\eq{
		\frac{dx_1}{dt} &= \kk_1 x_1 x_2 - \kk_2\kk_3 x_1^2 - \kk_3 x_1^3  \\
		\frac{dx_2}{dt} &= -\kk_1 x_1 x_2 + \kk_2\kk_3 x_1^2 + \kk_3 x_1^3
	}
corresponding to the reaction network $\rm{X}_1 + \rm{X}_2 \RR 2\rm{X}_1$ and $3\rm{X}_1 \FR 2\rm{X}_1 + \rm{X}_2$. To get from the original system to the mass-action system, we have multiplied the vector field by the non-zero scalar field $\kk_2 + x_1$; this preserves the \emph{trajectory curves} of the system and corresponds to a time-rescaling along the trajectories~\cite{Brunner_Craciun}.

\subsection{Stochastic Mass-Action Systems}
\label{sec:stochastic}

	For a system whose chemical species are in very low abundance, the notion of concentration may no longer be a meaningful quantity and {\em molecular count} should be used instead. In this scenario, the most common model is {\em stochastic mass-action kinetics}, where the dynamics is given by a continuous-time Markov process~\cite{Anderson_Kurtz}.
	
	There are  strong connections between  stochastic and deterministic mass-action systems~\cite{Kurtz_1972, Anderson_Craciun_Kurtz, Stochastic1, Stochastic2}. For example, under appropriate volume scaling, the solutions of the stochastic system converge to those of the deterministic system~\cite{Kurtz_1972}. 
Moreover, if the deterministic system is {\em complex-balanced}, then the stochastic system has a unique stationary distribution, which is a product of Poisson distributions~\cite{Anderson_Craciun_Kurtz}.

Alternatively, instead of studying a Markov process, one may  choose to study the time evolution of the distribution on the state space, as governed by the {\em chemical master equation}, a system of ordinary differential equations whose dimension is the size of the state space~\cite{Ingalls}.

\subsection{Reaction-Diffusion Equations}
\label{sec:RD}

If spatial inhomogeneity and specific diffusion rates play an important role, then one may use  partial differential equations (PDEs) to model that system. Reaction-diffusion equations are the most common such PDEs in practice. For example, they are used for analyzing biological pattern formation and, in particular, Turing patterns~\cite{Kondo_Miura_2010, Mincheva_Craciun}.  Recent work features strong connections between reaction-diffusion equations and complex-balanced mass-action systems~\cite{Fellner_Tang, Desvillettes_Fellner_Tang, Mohamed_Pantea_Tudorascu}.

%
%
%

\bibliographystyle{amsxport}
\bibliography{cite}


\end{document}